# Anomaly Detection in California Electricity Price Forecasting: Enhancing Accuracy and Reliability Using Principal Component Analysis

Joseph Nyangon, Ph.D.[1] *Senior Member*, IEEE and Ruth Akintunde, Ph.D.[2]
U.S. Department of Energy, Washington, DC., United States
[2] Research and Development Division, SAS Institute, Cary, NC, United States

*Abstract*— Accurate and reliable electricity price forecasting has significant practical implications for grid management, renewable energy integration, power system planning, and price volatility management. This study focuses on enhancing electricity price forecasting in California's grid, addressing challenges from complex generation data and heteroskedasticity. Utilizing principal component analysis (PCA), we analyze CAISO's hourly electricity prices and demand from 2016-2021 to improve day-ahead forecasting accuracy. Initially, we apply traditional outlier analysis with the interquartile range method, followed by robust PCA (RPCA) for more effective outlier elimination. This approach improves data symmetry and reduces skewness. We then construct multiple linear regression models using both raw and PCA-transformed features. The model with transformed features, refined through traditional and SAS Sparse Matrix outlier removal methods, shows superior forecasting performance. The SAS Sparse Matrix method, in particular, significantly enhances model accuracy. Our findings demonstrate that PCA-based methods are key in advancing electricity price forecasting, supporting renewable integration and grid management in day-ahead markets.

*Keywords — Electricity price forecasting, principal component analysis (PCA), power system planning, heteroskedasticity, renewable energy integration*

## I. Introduction

Accurate and reliable anomaly detection in California's Day-Ahead wholesale Electricity Market is pivotal, particularly when employing Principal Component Analysis (PCA) to navigate the inherent complexities of the market. The electricity market, characterized by its high volatility, nonlinearity, high frequency, mean reversion, and non-stationarity, presents significant forecasting challenges essential for grid management, renewable energy integration, and handling price fluctuations for optimal power system planning and operations [1], [2].

Advanced analytics, including extreme learning machine (ELM), kernel-based extreme learning machine (KELM), bivariate empirical mode decomposition (BED), and PCA, have been instrumental in addressing these challenges. For instance, Hong et al. [3] utilized a PCA neural network in conjunction with multi-layer feed-forward neural networks for day-ahead locational marginal price (LMP) forecasting in the PJM energy market. Similarly, Wang et al. [4] combined Grey Correlation Analysis (GCA) and Kernel function with PCA for feature extraction in price classification. In anomaly detection using kernel density estimation, the focus is on evaluating the kernel density function at a given test data point. The kernel density estimation function is defined as follows:

$$f_x(x) = \frac{1}{nh}\sum_{i=1}^{n} K\left(\frac{x - x_i}{h}\right)$$

Where the probability of a data point $x$ equals the sum of the kernel function at $x$, $x_i$ is a set of training data points, $i$ represents each individual data point in the training set, and $n$ is the total number of these points. The kernel function, represented as $K$, plays a crucial role in this estimation process. Additionally, the parameter $h$, known as the bandwidth, is crucial in determining the function's sensitivity. These methodologies underscore the utility of PCA in enhancing the accuracy and reliability of forecasts in complex electricity markets.

The California Independent System Operator (CAISO) Day-Ahead Market (DAM) has been a focal point of study due to its unique price behaviors, notably asymmetry and price-elasticity [5]–[7]. Negative prices in the California DAM, linked to renewable energy's merit-order effect, have garnered significant attention [8]. Filho et al. [9] and Byrne et al. [10] explored the implications of negative LMPs on grid operations and energy storage revenue optimization, respectively. These studies highlight the criticality of understanding price fluctuations for market stability and the development of renewable energy, transmission, and storage.

Moreover, the integration of renewable energy, influenced by government incentives like the Investment Tax Credit (ITC) and Production Tax Credit (PTC), has been a key factor in CAISO's market volatility. The COVID-19 pandemic further exacerbated this trend by reducing electricity demand, leading to negative prices in approximately 4% of all hours across U.S. wholesale market nodes [11]. This scenario underscores the importance of robust forecasting techniques in maintaining a stable and competitive market amidst evolving operational and management needs.

## II. Forecasting Renewable Energy Price Variations

### A. Electricity Price Forecasting

Electricity price forecasting is crucial for market participants in developing effective risk management strategies. This field is broadly categorized into five areas: statistical time series analysis, machine learning and data-driven techniques, econometric models, hybrid models, and behavioral finance.

Statistical methods, such as autoregressive integrated moving average (ARIMA), autoregressive (AR), autoregressive-exogenous (ARX(n, m)), Generalized AutoRegressive Conditional Heteroskedasticity (GARCH), exponential smoothing, and vector autoregression (VAR), analyze electricity prices' statistical behavior [1], [3], [7], [12], [13]. Machine learning, utilizing algorithms like PCA, Artificial Neural Networks (ANN), Support Vector Machine (SVM), decision trees, and random forests, forecasts prices based on historical data [14]. Econometric models examine economic factors influencing prices, such as weather and fuel costs [15]. Hybrid models blend multiple methodologies for enhanced accuracy, while behavioral finance focuses on human behavior's impact on pricing. Recent advancements show machine learning models, including relevance vector machine (RVM), deep learning, and semiparametric models, outperform traditional statistical approaches in forecasting accuracy [16], [17]. These methodologies, particularly statistical time series analysis and machine learning, are valuable in accurate and reliable electricity price forecasting, vital for informed decision-making in the power markets.

### B. Using PCA to Enhance Renewable Energy Integration

California's renewable energy integration has advanced significantly, driven by innovative methodologies like PCA. PCA's role in reducing data dimensionality and revealing patterns is pivotal. For instance, Van Triel and Lipman [18] demonstrated how PCA-enhanced tools like V2G-SIM and GridSim can predict electric vehicle (EV) charging impacts on renewable resources, potentially saving US$20 billion in storage investments by 2030. Similarly, Liu et al. [19] utilized PCA to extract key features of renewable energy, forming a basis for predictive generation models.

Further, Sreedharan et al. [20] emphasized the need for inventive incentives to support Distributed Energy Resources (DERs) in commercial and industrial sectors. Ginsberg et al. [21] used PCA in assessing costs of variable renewable energy integration, offering insights beyond California. Hopwood [22] applied PCA to current-voltage traces, enhancing failure classification accuracy. Nikkhah et al. [14] combined PCA with Artificial Neural Networks (ANN) for energy flow forecasting in agriculture, reducing data complexity. Talaat et al. [23] developed a dynamic model for hybrid renewable energy systems, improving efficiency. Vishnupriyan and Manoharan [24] used an Analytic Hierarchy Process (AHP) for optimal power system planning, focusing on a grid-connected PV system. These studies collectively highlight PCA's transformative impact in renewable energy integration, particularly in California's wholesale electricity market. By improving forecasting accuracy, reducing data complexity, and optimizing grid integration, PCA-driven approaches are crucial in shaping future renewable energy strategies.

### C. Day-ahead Price Forecasting and Net Load Analysis

The CAISO market system, headquartered in Folsom, is a pivotal non-profit entity managing California's high-voltage grid. It oversees three main market domains: ancillary services, real-time imbalance energy, and transmission congestion management. CAISO's Market Redesign and Technology Upgrade (MRTU), initiated on April 1, 2009, introduced a three-tiered day-ahead market process: market power mitigation, the integrated forward market, and the residual unit commitment process. The latter ensures the readiness of power plants for next-day generation. Within this framework, Scheduling Coordinators (SCs) play a pivotal role, linking the ISO with generators, retailers, and end-users, and influencing market prices through their bids. The system's efficiency hinges on a comprehensive network model that evaluates transmission and generation assets to determine the most cost-effective energy solutions, thereby informing SCs' transactions in the day-ahead market, which accepts bids up to a week in advance [25].

Day-ahead price forecasting, however, faces challenges in accurately predicting net load, leading to energy imbalances between forecasted and actual consumption. This discrepancy stems from the day-ahead market's hourly scheduling versus the real-time market's 15- and 5-minute intervals. Net load analysis, crucial in this context, calculates the difference between electricity demand and renewable energy supply. This process, while essential, has limitations, such as the "copperplate" model's exclusion of transmission and operational parameters. Despite these drawbacks, net load analysis remains critical for scenario exploration and informing capacity expansion and economic dispatch models [26]. CAISO's MRTU system and net load analysis are instrumental in enhancing the accuracy of day-ahead price forecasting. These tools not only facilitate efficient market operations but also improve the integration of renewable energy sources and grid stability maintenance.

## III. METHODOLOGY AND DATA EXPLORATORY ANALYSIS

### A. The PCA Methodology

PCA method is a pivotal technique in multivariate statistical analysis, extensively utilized in electricity market research and beyond [4], [19], [22]. It effectively transforms a set of related variables into a new set of uncorrelated variables, known as principal components (PCs), while retaining a variation in the original dataset. This transformation is achieved by identifying the variables that contribute most to the total variability in the original data set [3]. Each PC is a linear combination of the original variables, with the first PC accounting for the maximum variance, equivalent to the largest eigenvalue of the covariance matrix [27]. Subsequent PCs capture the remaining maximum variance not accounted for by the previous PCs. Importantly, each PC is uncorrelated with others, allowing for the selection of PCs with larger variance values for analysis [14], [22]. PCA's ability to reduce the dimensionality of large data sets is valuable in power market studies [1], [4], [22]. It reduces dimensionality of variables into fewer PCs, thus summarizing a big portion of the data's variation. This technique uncovers latent common structures and interprets the structural meaning of each PC. The efficiency and reliability of PCA makes it an indispensable tool in power market analysis.

### B. Exploratory Analysis and Covariates

Table 1 summarizes electricity prices for 24 hours each day and 365 days each year. This translates to 52441 observations. Our study, spanning 2016 to 2021, examined electricity prices and demand, revealing significant fluctuations in prices. These variations are aligned with specific socio-economic events.

TABLE I. DESCRIPTIVE STATISTICS OF ELECTRICITY PRICES FROM 2016 TO 2021

|       | 2016  | 2017   | 2018   | 2019   | 2020   | 2021  |
|-------|-------|--------|--------|--------|--------|-------|
| Count | 8760  | 8665   | 8736   | 8760   | 8784   | 8736  |
| Mean  | 29.85 | 34.90  | 39.48  | 37.12  | 33.42  | 53.92 |
| Std   | 10.77 | 27.39  | 32.13  | 22.86  | 34.82  | 37.99 |
| Min   | 0.35  | -13.23 | -17.33 | -12.59 | -10.43 | -0.31 |
| Max   | 103.3 | 806.3  | 946.4  | 259.3  | 997.4  | 961   |

Notably, sharp price spikes, exceeding 200 cents/kWh, occurred intermittently from 2017 to 2021. However, these spikes are not solely attributable to demand fluctuations. Our correlation analysis, after outlier adjustment using the Sparse Matrix method, indicates a low to moderate positive correlation between price and demand for 2019 (rho=0.296) and 2021 (rho=0.512), respectively. This implies that factors beyond demand significantly influence electricity prices. California's wildfires have notably impacted electricity markets, both economically and regulatory. These fires, causing extensive damage to transmission lines, have forced utilities to undertake costly infrastructure upgrades, leading to higher consumer rates. The bankruptcy of PG&E in 2018, with over $30 billion in liabilities due to wildfires [28], exemplifies this impact and has prompted state intervention proposals. Wang et al. [29] estimate that wildfires have cost approximately 1.5% of California's GDP in indirect losses. Furthermore, Malloy [30] highlights that liability regulations have spurred utilities to enhance safety measures [31] discusses the health implications of PG&E's public safety power shutoffs, designed to reduce wildfire risks. These factors collectively underscore the complex interplay between natural disasters, regulations, and market dynamics.

### C. PCA and SAS Robust PCA Anomaly Detection Analysis

We employed two distinct methods to identify and remove outliers in the electricity prices, enhancing the accuracy of our data. Initially, we utilized the Interquartile Range (IQR) method, defining outliers as values falling outside the bounds of 1.5 times the IQR from the first and third quartiles (Q1 and Q3). This approach is visually represented in Figure 1, which displays a boxplot of electricity prices for every hour of each day from 2016 to 2021, prior to outlier removal. Post-removal, Figure 2 illustrates the refined distribution using this traditional method.

Further refining our analysis, we implemented the Robust Principal Component Analysis (RPCA) technique using SAS Viya platform. This advanced method, detailed by Salis et al. [32], decomposes the data matrix into a low-rank matrix and a sparse matrix, the latter containing outliers, as follows:

$$\text{minimize } \| L \|_* + \lambda \| S \|_1$$
$$\text{subject to } M = L + S$$

Where the nuclear norm is defined as the nuclear norm for $L$, $\| S \|_1$ is the norm of matrix $S$. The parameter $\lambda$ plays a crucial role in balancing these two terms. A higher value of $\lambda$ results in a sparser matrix $S$, while a lower $\lambda$ leads to a lower rank for matrix $L$. Zhao et al. [33] demonstrated that under certain conditions, such as the low-rank component not being too large and the sparse component being sufficiently sparse, setting $\lambda$ to:

$$\lambda = \frac{1}{\sqrt{n}}$$

where $n$ is the number of observations in matrix $M$.

Our focus was on daily electricity prices, using the 'Day of Year' (DOY) feature for anomaly detection. This approach, supported by Hong & Wu [3], who considered variables like 'weekday' and 'local demand' for forecasting, proved effective in our study. The resulting sparse matrix, named *sparsemat2*, as outlined in [34], identified outliers, which were then excluded from our analysis. Figures 3 showcases the outcomes of outlier removal using the SAS Sparse Matrix strategy alone and in combination with the IQR method, respectively.

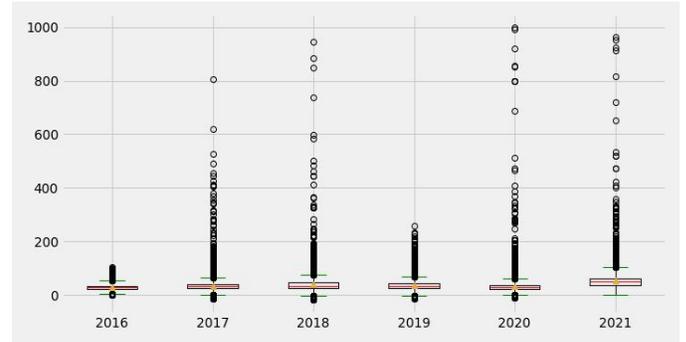

Fig. 1. Boxplot of hourly electricity prices from 2016 to 2021

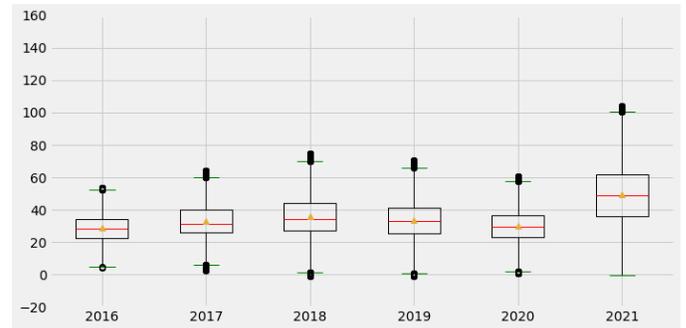

Fig. 2. Boxplot after outlier removal using the traditional strategy (1.5*IQR)

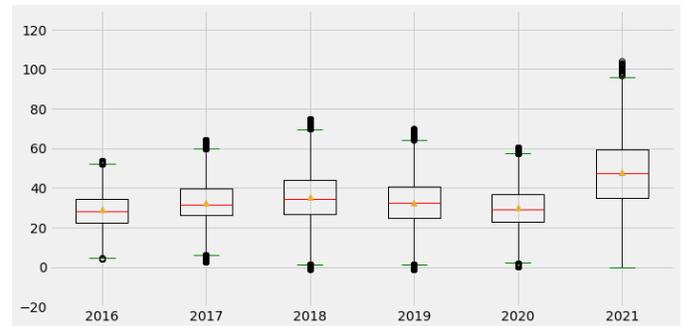

Fig. 3. Comparison of boxplots with outlier removal using traditional and SAS' Sparse Matrix strategy

This dual-method approach to outlier detection and removal in electricity price ensures a robust and accurate analysis, crucial for informed decision-making in the power markets.

## IV. RESULTS AND DISCUSSION

### A. Results

The raw features used for hourly electricity price prediction include *YDay Price* (yesterday's price for that hour), *YDay_Load* (yesterday's load for that hour), *YAve_Load* ( yesterday's average load for that hour), Month (the numeric month of the year), *DOW* (the numeric day of the week), DOM (the numeric day of the month) and *DOY* (the day of the year, values are between 1-366). We developed SAS models for electricity market analysis and explored PROC PCA for PC generation for regression analysis and SAS' PROC RPCA anomaly detection feature for outlier removal analysis (see Table II). We examined up to 5 PCA features for predictions and applied the PROC RPCA anomaly detection, as described by [34].

Model 3, the linear regression model using raw features after outlier removal through both the traditional and SAS Sparse Matrix methods, achieved the highest performance on the training set. This shows that the SAS Sparse Matrix outlier removal method, implemented via PROC RPCA enhanced model performance compared to Model 2 where only the traditional outlier removal method was applied. Model 4, which employed SAS PCA features after outlier removal using traditional and SAS Sparse Matrix methods, achieved the highest performance on the test set: a slight improvement when SAS PCA features are applied instead of the 7 raw features.

TABLE II. RESULTS SHOWING PERFORMANCE OF SELECTED MODELS

| S/N | Model Method | Training RMSE | Test RMSE | Training $R^2$ | Test $R^2$ |
|---|---|---|---|---|---|
| 1 | Model using 7 raw features before outlier removal | 18.65 | 26.28 | 0.63 | 0.18 |
| 2 | Model using 7 raw features after outlier removal by traditional method (1.5*IQR w/ Raw) | 6.19 | 6.69 | 0.84 | 0.82 |
| 3 | Model using 7 raw features after outlier removal by traditional then SAS Sparse Matrix method | 5.72 | 5.87 | 0.85 | 0.84 |
| 4 | Model using SAS PCA transformed features after outlier removal by traditional then SAS Sparse Matrix method | 5.98 | 5.83 | 0.83 | 0.84 |

### B. Discussion

To enhance forecasting accuracy, our study focused on combining outlier removal methods, like the SAS' Sparse Matrix method, with feature transformations such as PCA. Outlier removal methods effectively eliminate extreme data values, reducing the impact of anomalies and noise on the forecasting model. PCA, a technique for reducing data dimensionality, identifies key variables that contribute most to data variability. This method has been effective in electricity market analyses, helping to capture price volatility dynamics and fluctuations. Applying PCA to input variables captured underlying data patterns, crucial for accurate and reliable price forecasting. This approach is particularly relevant given the significant price fluctuations influenced by events like California wildfires, which impact electricity transmission infrastructure and lead to higher consumer rates. We showed that integrating outlier removal and PCA improves day-ahead electricity price forecasting models in the CAISO market. This combination reduces the complexities of variable renewable energy integration, improves market efficiency.

## V. CONCLUSION

By integrating outlier removal techniques, feature transformations like PCA, and supervised learning methods, this study demonstrates a significant improvement in managing heteroskedastic noise, which is pivotal in power market dynamics, like CAISO. These methods helps adapt to changes in the generation mix, policy uncertainties, and integration of renewable electricity like solar PV and wind power [35]–[38].

The implications of this research extend to policymakers, utility operators, and grid managers, highlighting the importance of technological innovation in forecasting models. Incorporating PCA and supervised learning enhances the speed and accuracy of predictions, supporting effective decision-making and resource planning in a market heavily reliant on renewable energy [39], [40]. The study also underlines the limitations of net load analysis and the need to consider external factors like wildfires, seasonal trends, and temperature changes, as well as transmission, operational parameters, and stability metrics, to improve forecasting model accuracy.

ACKNOWLEDGMENT

The authors acknowledge the SAS Institute for computing and providing the SAS Viya 4 simulation platform. We also acknowledge valuable inputs from anonymous reviewers, which significantly improved this research.